\documentclass[amsmath,trackchanges]{aastex702}
\usepackage{subcaption}

\usepackage{longtable}
\usepackage{booktabs}
\shorttitle{Observational evidence of black hole event horizon}
\shortauthors{Ma et al.}

\begin{document}

\title{Observational evidence of black hole event horizon from ultra-compact X-ray binaries}
	
\author{Qian-Qi Ma}
\affiliation{Department of Astronomy, Xiamen University, Xiamen, Fujian 361005, People’s Republic of China}
\email{maqianqi1@stu.xmu.edu.cn}

\author{Wei-Min Gu}
\affiliation{Department of Astronomy, Xiamen University, Xiamen, Fujian 361005, People’s Republic of China}
\email[show]{guwm@xmu.edu.cn}
\correspondingauthor{Wei-Min Gu}

\author{Shan-Shan Weng}
\affiliation{School of Physics and Technology, Nanjing Normal University, Nanjing, 210023, People’s Republic of China}
\email[show]{wengss@njnu.edu.cn}
\correspondingauthor{Shan-Shan Weng}

\author{Mouyuan Sun}
\affiliation{Department of Astronomy, Xiamen University, Xiamen, Fujian 361005, People’s Republic of China}
\email{msun88@xmu.edu.cn}

\begin{abstract}
\noindent 
The event horizon is the most crucial feature that distinguishes black hole (BH) from all other objects in the universe. However, since no radiation can escape from the event horizon, its direct observational evidence remains lacking. Although both are compact objects, the event horizon leads to differences in the radiative behavior between black hole and neutron star (NS) X-ray binaries. Historically efforts to probe the existence of event horizons revealed that the BH X-ray binaries are fainter than NSs with similar orbital periods in their quiescent states. Here, we use a special class of X-ray binaries namely the ultra-compact X-ray binaries (UCXBs), to reveal the existence of the event horizon near stellar-mass black holes. In short-period UCXBs, gravitational radiation dominates the orbital angular momentum loss, allowing us to accurately determine their accretion rates. Based on five UCXBs with orbital periods shorter than 40 min, located in globular clusters, and identified accretors (NS or BH), we obtained their accretion rates. Combining their accretion rates and X-ray luminosities, we calculated the X-ray radiative efficiencies of NS and BH. The radiative efficiency of BH is about two to three orders of magnitude lower than that of NS systems, indicating the accretion energy is swallowed by the BH, thereby indirectly probe the presence of the event horizon.
		
\end{abstract}
	
\keywords{}
	
\section{Introduction}

The observations of the central black hole (BH) in M87 and Sagittarius A* (Sgr A*) by the Event Horizon Telescope (EHT) have directly imaged the black hole shadow \citep{2019ApJ...875L...1E, 2022ApJ...930L..12E}, which shows a characteristic consequence of photons being captured near the unstable circular orbit and provide indirect evidence for the existence of the event horizon. The central BH in M87 and Sgr A* are both supermassive black holes, whose shadows have comparable angular scales of the resolution of EHT \citep[e.g.,][]{2019ApJ...875L...2E, 2022ApJ...930L..12E}. However, since the event horizon radii of stellar-mass BHs are only tens of kilometers and their shadows are significantly smaller than the angular resolution of EHT, they cannot be directly imaged. 

\begin{figure}[h]
    \centering
    \includegraphics[width=0.8\textwidth]{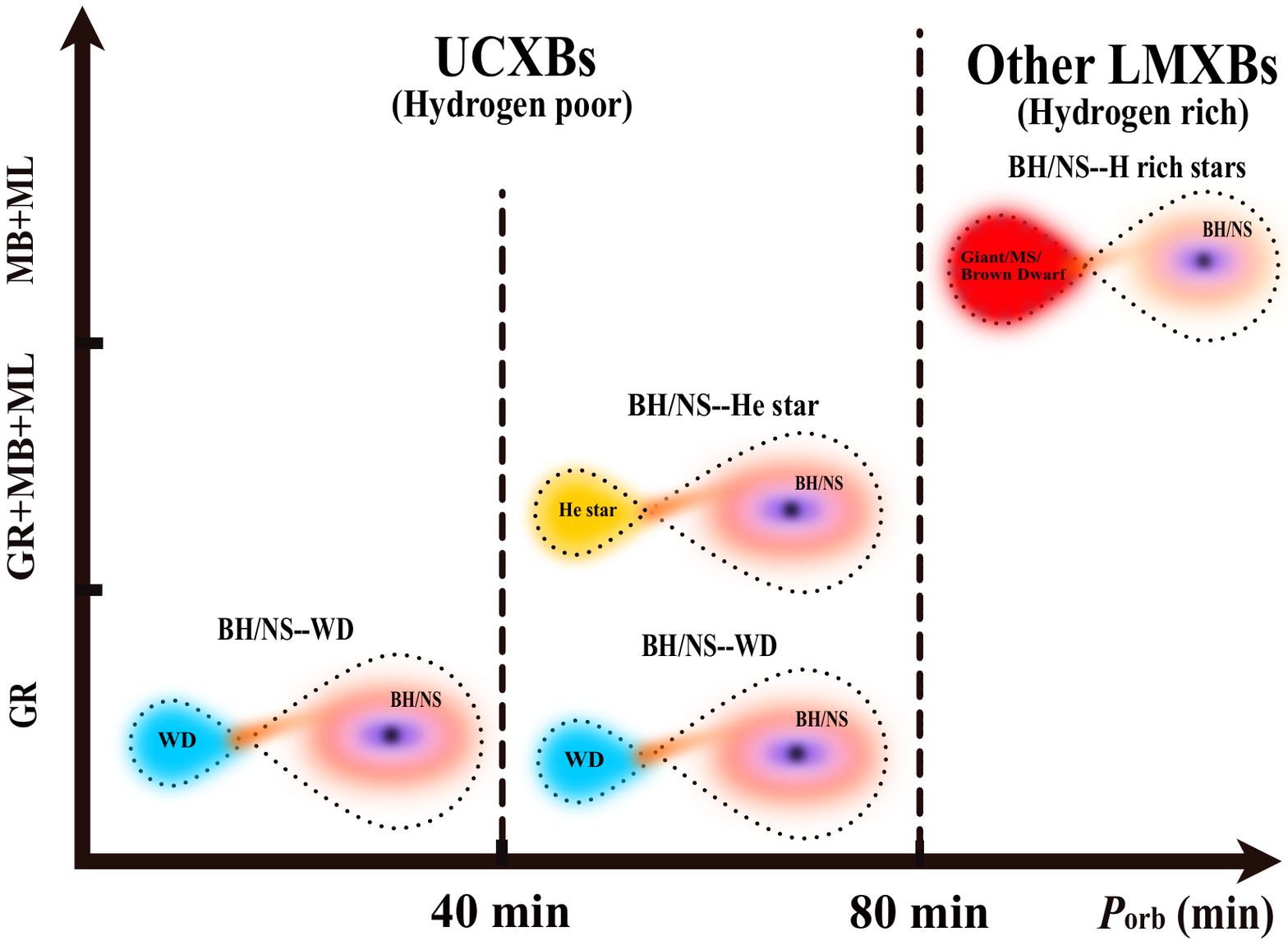}
    \caption{The dominated orbital angular momentum loss mechanisms in LMXBs with different $ P_{\rm orb} $. For the systems with WD companions, GR dominates the orbital angular momentum loss and the accretion rate can be accurately calculated; for systems with other types of companion stars, orbital angular momentum loss results from multiple mechanisms include GR, MB, and ML.}
    \label{fig:1}
\end{figure}

For stellar-mass BHs in low-mass X-ray binaries (LMXBs), the quiescent luminosities are usually two orders of magnitude lower than those of neutron star (NS) systems for similar orbital periods \citep[e.g.,][]{2001ApJ...553L..47G, 2004ApJ...615..402M, 2008NewAR..51..733N}, which often interpreted as the indirect evidence that the BH event horizon swallows most of the accreting energy. This method is mainly applied to LMXBs with orbital periods ($ P_{\rm orb} $) longer than 80 min. In these LMXBs, the complexity of the hydrogen rich companion star such as brown dwarf \citep{2001ApJ...557..292B}, main-sequence star \citep[e.g.,][]{1998ApJ...499..375O, 2002MNRAS.334..233T, 2010A&A...516A..58G}, and giant \citep[e.g.,][]{1992Natur.355..614C, 1997ApJ...477..876O} results in multiple possible mechanisms for orbital angular momentum loss, such as magnetic braking (MB) \citep[e.g.,][]{1983ApJ...275..713R, 2013ApJ...768..184H} and the mass loss (ML) \citep[e.g.,][]{2008NewAR..51..860Y, 2010ApJ...717..724G, 2026arXiv260921324T}, making it difficult to determine the accretion rate (see Figure \ref{fig:1}). Therefore, the conclusion relies on the assumption that for both BH and NS systems, similar orbital periods correspond to comparable accretion rates \citep{2001ApJ...553L..47G}.

Ultra-compact X-ray binaries (UCXBs) are a sub-population of LMXBs, which exhibit no hydrogen lines in the optical spectrum and have orbital periods shorter than 80 min \citep{1981ApJ...248L..27P}. For UCXBs with $ 40 \ \rm min < P_{\rm orb} < 80 \ min $, the companion stars are white dwarfs (WDs) or helium stars (He stars) and their angular momentum losing mechanisms are uncertain. For short-period UCXBs with $ P_{\rm orb} < 40 $ min, the tight orbital separation only allows a WD companion \citep[e.g.,][]{2013ApJ...768..184H, 2021MNRAS.506.4654W} (Figure \ref{fig:1}). Gravitational radiation (GR) dominates the orbital angular momentum loss and drives mass transfer \citep[e.g.,][]{2013ApJ...768..184H, Ma2026ExtragalacticUCXB}, and their accretion rate depends only on gravitational theory and the equation of state of WDs. Thus, we can derive ``clean" accretion rates from them, and the radiative efficiencies obtained in these short-period UCXBs are robust. 

In this work, we derived the radiative efficiencies of five short-period UCXBs from their accretion rates and long‑term average X‑ray luminosities. Our results show that the radiative efficiency of BH is significantly lower than that of NS systems, which constitute a strong evidence for the existence of BH event horizon.

\section{Methods} \label{sec:2}
We focus on short-period UCXBs with NSs or BHs that satisfy the following criteria:

(1) Accurate distances: all sources are located in globular clusters (GCs), to ensure reliable distance and luminosity measurements;

(2) WD companions: the orbital periods of all sources are shorter than 40 min, to ensure their companion stars are WDs and GR dominates the angular momentum loss;

(3) Accretors (NS or BH) have been identified: NSs are all identified through Type I X-ray bursts \citep[e.g.,][]{1976ApJ...207L.177F, 1993MNRAS.262..511E, 2016PASJ...68...95S, 2025A&A...695A..44D}, and the BH is identified through the radio/X-ray correlation in the hard state \citep{2015MNRAS.453.3918M}. We obtained five UCXBs, whose basic properties are listed in Table~\ref{tab:1}, and detailed observational information for these sources is provided in Appendix, Tables \ref{tab:2}-\ref{tab:6}.

\begin{deluxetable*}{lllll}[]
    \tablenum{1}
    \tablewidth{0pt}
    \tablecaption{The properties of five selected UCXBs \label{tab:1}}
    \tablehead{
        \colhead{Object} & \colhead{$ P_{\rm orb} $ (s)\tablenotemark{1}} & \colhead{Type} & $ d $ (kpc)\tablenotemark{2} & $ \overline{L}_{\rm X}\:(\rm erg\,s^{-1}) $\tablenotemark{3} \\
    }
    \startdata
    4U 1820-303 & 685 & NS-WD & 8.0 & $ 6.4 \times 10^{37} $\\
    4U 0513-40  & 1020 & NS-WD & 12.0 & $ 5.5 \times 10^{36} $\\
    4U 1850-087 & 1236 & NS-WD & 7.4 & $ 2.4 \times 10^{36} $\\
    M15 X-2 & 1356 & NS-WD & 10.7 & $ 3.9 \times 10^{36} $ \\
    47 Tuc X-9 & 1692 & BH-WD & 4.5 & $ 3.5 \times 10^{33} $ \\
    \enddata
    \tablecomments{The parameters listed in the table include the orbital period ($ P_{\rm orb} $), the type of the system (Type), the distance ($ d $), and the time-averaged X-ray luminosity ($ \overline{L}_{\rm X} $).}
    \tablenotetext{1}{4U 1820-303: from \cite{1987ApJ...312L..17S}; 4U 0513-40: from \cite{2009ApJ...699.1113Z}; 4U 1850-087: from \cite{1996MNRAS.282L..37H}; M15 X-2: from \cite{2005ApJ...634L.105D}; 47 Tuc X-9: from \cite{2017MNRAS.467.2199B}.}
    \tablenotetext{2}{All distances are from \cite{2021MNRAS.505.5957B}, with uncertainties below 2\%.}
    \tablenotetext{3}{The X-ray energy band is 0.5-10 keV.}
\end{deluxetable*}

For BH/NS-WD accreting systems, the orbital angular momentum loss is dominated by the GR, which drives the WD donor to continuously fill its Roche lobe and sustain mass transfer \citep[e.g.,][]{2018MNRAS.475L.101D, 2020MNRAS.497.1543G, Ma2026ExtragalacticUCXB}. We derive the accretion rates by only adopting the Roche lobe theory, the mass-radius relation of WD. For the WD to fully fill the Roche lobe, the WD mass ($M_2$) should be \citep{2020MNRAS.497.1543G}
\begin{equation}
    M_{2} = 0.1 M_{\odot} \left(\frac{P_{\rm orb}}{471 \rm \: s}\right)^{-1}, \label{eq:1}
\end{equation}
and the accretion rate is 
\begin{equation}
    \dot{M} = 2.11 \times 10^{18} \left(\frac{2}{3} - \frac{M_{2}}{M_{1}}\right)^{-1} \left(\frac{M_{1}}{M_{\odot}}\right) \left(\frac{M_{2}}{0.1 M_{\odot}}\right)^{\frac{14}{3}} \left(\frac{M_{1} + M_{2}}{M_{\odot}}\right)^{-\frac{1}{3}} \ \rm g \, s^{-1}, \label{eq:2}
\end{equation}
where $ M_{1} $ is the mass of the compact object and $ M_{\odot} $ is the solar mass, and we assume that $ M_{1} $ for NS is in the range of 1.4-2 $ M_{\odot} $, while for BH is 3-20 $ M_{\odot} $. The detailed derivations of Equations (\ref{eq:1}) and (\ref{eq:2}) are in Appendix.

\section{Results} \label{sec:3}

\begin{figure}[h]
    \centering
    \includegraphics[width=0.8\textwidth]{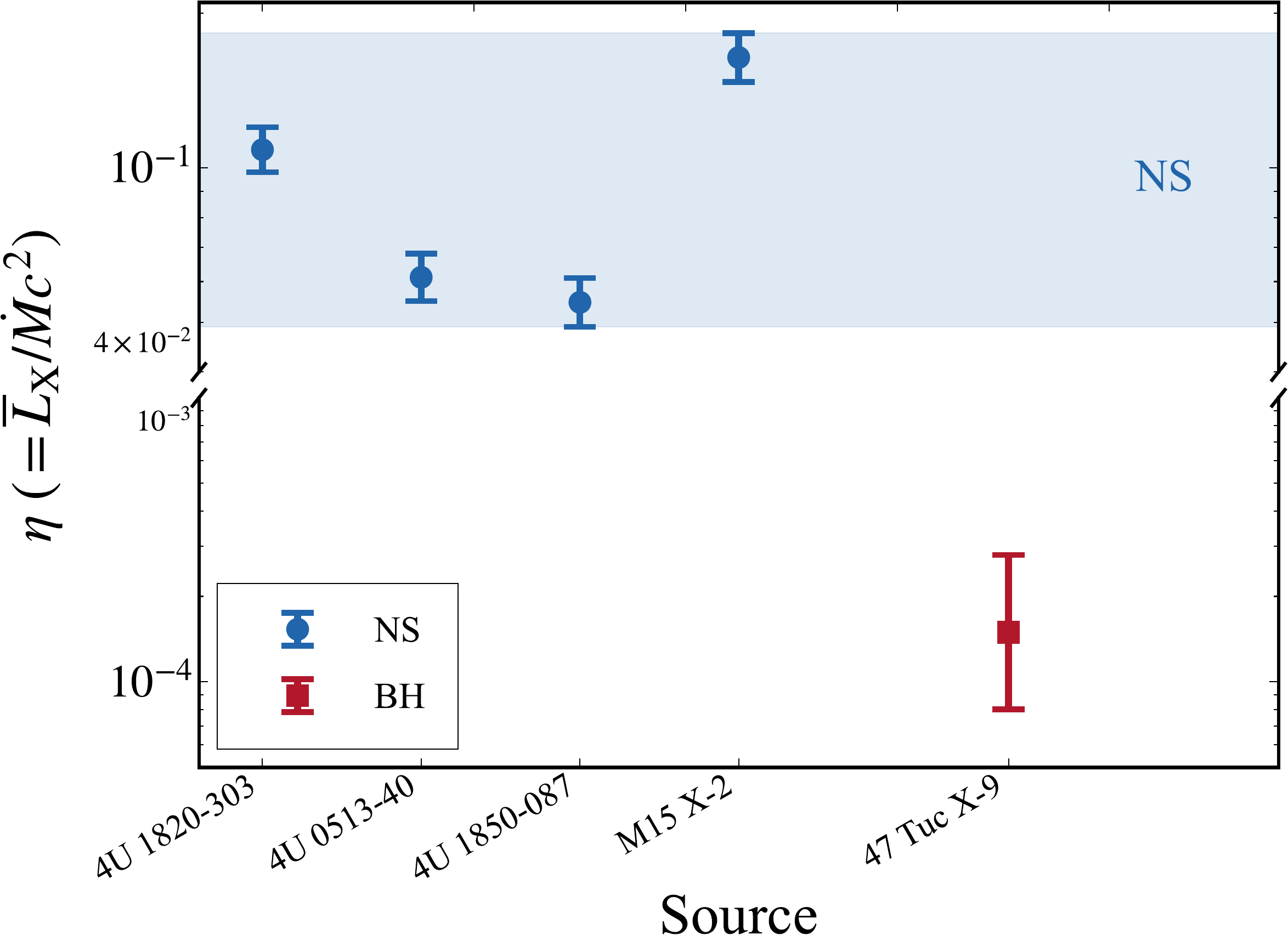}
    \caption{The radiative efficiencies ($ \eta $) of the five UCXBs. The uncertainties of $ \eta $ come from the mass values, where we assume the mass of NS is in the range of 1.4-2 $ M_{\odot} $, and for BH is 3-20 $ M_{\odot} $. The radiative efficiencies for all NSs are higher than 0.05, while for BH is lower than $ 3 \times 10^{-4} $. Therefore, the radiative efficiency of BH system is two to three orders of magnitude lower than that of NS system.}
    \label{fig:2}
\end{figure}

Based on the time-averaged X-ray luminosity ($ \overline{L}_{\rm X} $ in Table \ref{tab:1}) and the accretion rate ($\dot{M}$) from Equation (\ref{eq:2}), the radiative efficiencies for the five UCXBs can be derived as
\begin{equation}
    \eta = \overline{L}_{\rm X} / (\dot{M} c^{2}). \label{eq:3}
\end{equation}
The result of $ \eta $ is shown in Figure \ref{fig:2}. It is evident that the radiative efficiency of BH is lower than that of NS by two to three orders of magnitude.

\begin{figure}[h]
    \centering
    \includegraphics[width=0.8\textwidth]{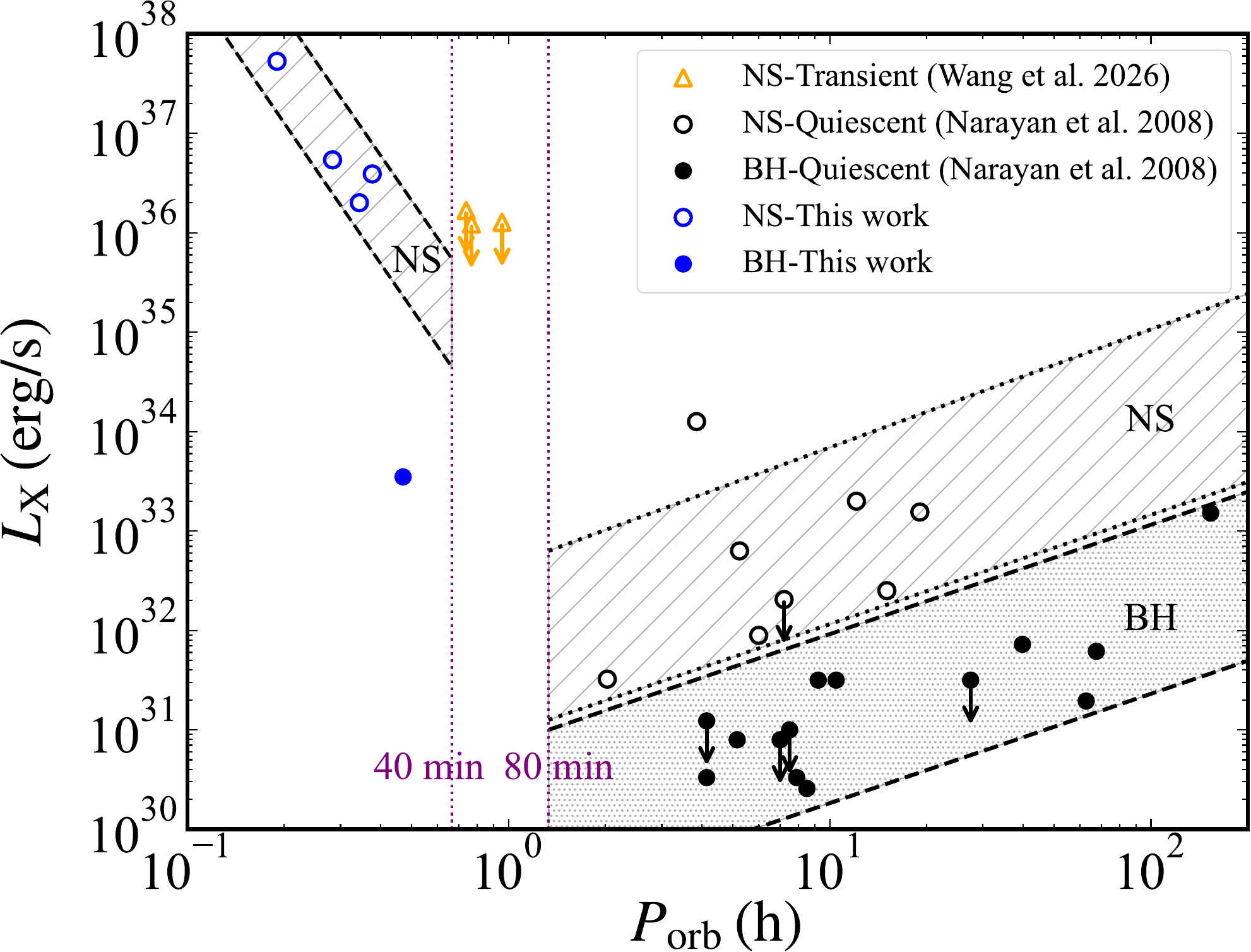}
    \caption{The known UCXBs in GCs ($ P_{\rm orb} < 80 $ min) and other LMXBs ($ P_{\rm orb} > 80 $ min). Blue hollow and solid circles: UCXBs in this work, the time-averaged X-ray luminosities in the 0.5-10 keV band; yellow triangles: transient UCXBs in GCs, peak X-ray luminosities in the 2-10 keV band, from \cite{2026RAA....26k2001W}; black hollow and solid circles: LMXBs in the quiescent state from \cite{2008NewAR..51..733N}, the lowest observed X-ray luminosities or the upper limits derived from \textit{Chandra}/\textit{XMM} observations in the 0.5-10 keV band.}
    \label{fig:3}
\end{figure}

The luminosities comparison of UCXBs to other LMXBs is shown in Figure \ref{fig:3}. In this figure, there are three separated regions, which include: (i) $ P_{\rm orb} < 40 $ min (blue hollow and solid circles), where the selected five UCXBs in this work have WD companions; (ii) $ 40 \: \mathrm{min} < P_{\rm orb} < 80 \: \mathrm{min} $ (yellow triangles), for which the companion types of those UCXBs are unknown and are transient sources, so we only use their peak outburst luminosities \citep[e.g.,][]{2023A&A...677A.186A, 2026RAA....26k2001W}; (iii) $ P_{\rm orb} > 80 $ min (black hollow and solid circles), where other LMXBs are all in their lowest observed quiescent states, taken from Fig.~9 of \cite{2008NewAR..51..733N}. All these sources should accrete through Roche lobe overflow, but with different mechanisms driving the mass transfer. As aforementioned, accretion rate can only be reliably calculated for short-period UCXBs with $P_{\mathrm{orb}}<40\ \mathrm{min}$. For the other two regions, their accretion rates may positively or negatively correlated with the orbital period since their angular momentum transfer processes are mixed. Nevertheless, the accretion rates of these LMXBs may be much smaller than those of short-period UXCBs and thus have lower X-ray luminosities. Unlike, short-period UCXBs, the accretion rates of these LMXBs are unknown and the corresponding radiative efficiency cannot be calculated.

\section{Conclusions and Discussion}  \label{sec:4}

The accretion process convert the released gravitational energy into thermal, kinematics, or radiation energy. For gas accreted from infinity, the gravitational potential is $ -G M / R_{\ast} $, where $ R_{\ast} $ is the radius of the central object. If the accretor is an NS, which has a hard surface, the observed luminosity can be divided into two components: emission from the accretion flow ($ L_{\rm flow} $) and from the interaction between accreting matter and the surface of NS ($ L_{\rm surf} $). In a steady accretion flow in which most of the transferred mass ultimately reaches the surface of an NS, the total luminosity is approximately $ L_{\rm tot} = L_{\rm flow} + L_{\rm surf} \approx G M \dot{M} / R_{\ast} $. Even if the accretion flow is radiatively inefficient and advects a substantial fraction of the dissipated energy inward, this energy must ultimately be deposited at the surface of NS and can be reradiated \citep{2008NewAR..51..733N}. According to recent research, an NS with $ M = 1.4 \ M_{\odot} $ has a radius of $ R_{\ast} \sim 12 $ km \citep{2025PhRvD.111c4005B}, and its radiative efficiency is $ \eta = L_{\rm tot} / \dot{M} c^{2} \sim 0.17 $, which is approximately independent of $ \dot{M} $ \citep{1995ApJ...452..710N} and roughly consistent with our calculations (Figure \ref{fig:2}).

For the BH, by contrast, there is no material surface absorbs all advected gas energy. In the radiatively efficient standard thin disk regime, the luminosity is determined by the binding energy released before the gas reaches the innermost stable circular orbit (ISCO), and the observed luminosity only comes from the contribution of the accretion flow, i.e., $ L_{\rm tot} = L_{\rm flow} $ \citep{1997ApJ...478L..79N}. Therefore, the luminosity is $ L_{\rm flow} \approx G M \dot{M} / 2R_{\rm ISCO} $ for radiatively efficient accretion flow, where $ R_{\rm ISCO} $ is the radius of the ISCO \citep[e.g.,][]{2008NewAR..51..733N, 2008bhad.book.....K}. Assuming a non-spinning Schwarzschild BH with $ R_{\rm ISCO} = 6 G M / c^{2} $, the radiative efficiency is $ \eta = L_{\rm flow} / \dot{M} c^{2} \sim 0.1 $. However, if $ \dot{M} $ is lower than 1\% Eddington accretion rate, the inner accretion flow will transition to a radiatively inefficient accretion flow, such as an advection-dominated accretion flow (ADAF) \citep[e.g.,][]{1995ApJ...452..710N, 1997ApJ...489..865E, 2008NewAR..51..733N}. In this regime, a substantial fraction of the dissipated energy is stored as thermal energy and advected inward with the accreting plasma, and the radiative efficiency decreases with decreasing $\dot{M}$ \citep[e.g.,][]{1995ApJ...452..710N,2008NewAR..51..733N}. If a BH accretes via an ADAF, the observed luminosity is $ L_{\rm tot} = L_{\rm flow} \ll G M \dot{M} / 2R_{\rm ISCO} $ and $ \eta \ll 0.1 $ \citep{2008NewAR..51..733N}. At low accretion rates, both BH and NS may host radiatively inefficient hot accretion flows. Unlike an NS, the advected thermal energy in a BH system is swallowed by the event horizon and the total radiative efficiency $\eta \ll 0.1$ \citep{1995ApJ...452..710N}.

Based on the above discussion, for BH systems, the accretion can be either radiatively efficient or radiatively inefficient. But for NS systems, although the accretion flow of an NS may be radiatively inefficient, the presence of its hard surface prevents its total radiative efficiency from becoming low \citep{1995ApJ...452..710N}. Therefore, for 47 Tuc X-9, if it has a hard surface similar to that of an NS, its radiative efficiency should fall within the NS region in Figure \ref{fig:2}. For BH-WD systems, the boundary between radiatively efficient and inefficient lies at $ P_{\rm orb} \sim 20 $ min \citep{Ma2026ExtragalacticUCXB}, while NS systems remain radiatively efficient. Therefore, Figure \ref{fig:3} can also be used to make predictions for future newly discovered UCXBs with $ P_{\rm orb} < 40 $ min: all NS systems should lie in the shaded region, while BH systems with $ P_{\rm orb} $ close to 47 Tuc X-9 should have luminosities similar to it.

In this work, we derive that the accretion rates of M15 X-2 and 47 Tuc X-9 are similar, both on the order of $ 10^{16} \rm \ g \, s^{-1} $, but the luminosity of 47 Tuc X-9 is three orders of magnitude lower than M15 X-2. This result also has a broader implication for the interpretation of the growth of supermassive black holes (SMBHs): luminosity is not an appropriate proxy for $ \dot{M} $ in BHs \citep{1978MNRAS.184...53B}, and the low radiation luminosity does not necessarily imply the correspondingly low rate of mass supply to the BH \citep{2026arXiv260621105Z}. A recent analysis of the origins and growth of SMBHs likewise argues that the early growth of SMBHs may involve either super-Eddington accretion or substantially reduced radiative efficiencies, rather than requiring all systems to grow through standard radiatively efficient accretion with $ \eta \sim 0.1 $ \citep{2025MNRAS.536.3177A}. High-Eddington systems can become radiatively inefficient because of photon trapping and advection in slim disks \citep{2025NewAR.10101733A}. Hence, radiatively inefficient accretion can hide a significant fraction of the mass-growth process from luminosity based surveys at both very low and super-Eddington accretion rates.
	
\begin{acknowledgments}

We thank Bei You for helpful discussion.

\end{acknowledgments}

\appendix
\section{Numerical calculation}

The BH/NS–WD binary system is assumed to have a circular orbit, which follows Kepler’s third law ($ GM / a^{3} = 4 \pi^{2} / P_{\rm orb}^{2} $, where $ a $ is the orbital separation). The Roche lobe radius ($ R_{\rm L2} $) of WD is
\begin{equation}
    \frac{R_{\rm L2}}{a} = 0.462 \left(\frac{M_{2}}{M_{1} + M_{2}}\right)^{\frac{1}{3}}, \label{eq:A1}
\end{equation}
and the WD's mass-radius relation is \citep{1997MNRAS.291..732T}
\begin{equation}
    \frac{R_{2}}{R_{\odot}} = 0.0115 \sqrt{\left(\frac{1.44 M_{\odot}}{M_{2}}\right)^{2/3} - \left(\frac{M_{2}}{1.44 M_{\odot}}\right)^{2/3}}, \label{eq:A2}
\end{equation}
where $ R_{2} $, $ R_{\odot} $ are the WD radius and solar radius, respectively. For the UCXBs mentioned in this work, $ M_{2} < 0.1 M_{\odot} $, therefore Equation (\ref{eq:A2}) can be simplified as $ R_{2} / R_{\odot} = 0.0115 (M_{2} / 1.44 M_{\odot})^{-1/3} $.

The system accretes through Roche lobe overflow, thus $ R_{\rm L2} = R_{2} $, and combining Kepler’s third law, Equations (\ref{eq:A1}) and simplified Equation (\ref{eq:A2}), we obtain the relation between $ M_{2} $ and $ P_{\rm orb} $, i.e. Equation (\ref{eq:1}). The five selected UCXBs are all persistent X-ray sources, and thus their accretion are also stable. To maintain the accretion, WD always fills its Roche lobe, thus $ \dot{R}_{\rm L2} = \dot{R}_{2} $. Employing $ \dot{R}_{\rm L2} = \dot{R}_{2} $, the binary separation expansion rate (see Equation 4.14 of \cite{2002apa..book.....F}), and the orbital angular momentum loss rate by GR (see Equation 5.5 of \cite{1964PhRv..136.1224P}), the accretion rate $ \dot{M} $ can be derived, i.e. Equation (\ref{eq:2}). 

\section{Detailed observational information of the five UCXBs}

Generally, X-ray sources in GCs are so dense that observations with high spatial resolution are often required. For the UCXBs in this work, the \textit{Chandra} images of M15 X-2 \citep{2005MNRAS.357..325H} and 47 Tuc X-9 \citep{2017MNRAS.467.2199B} reveal the presence of several contaminating sources of comparable brightness around them. Therefore, only \textit{Chandra} data are used for their light curves. However, for 4U 1820-303, 4U 0513-40 and 4U 1850-087, these sources are the brightest within a 6-arcmin field of view (see their host GCs' images from \textit{Swift} LSXPS catalogue), so contamination from nearby sources is negligible, and data from other observatories can also be used. Detailed observational information for these sources are discussed in Appendix.

Accurate distances are crucial for determining the absolute X-ray luminosities of X‑ray binaries. Their distance measurements mainly rely on the following methods: (i) the host GC, for which the distance is often determined by combining parallax and the absolute luminosities of multiple stars, with an uncertainty of about 2\% \citep{2021MNRAS.505.5957B}; (ii) the absolute luminosity of the optical companion in quiescence, which yields a distance uncertainty of about 10\% and also provides a relatively reliable distance estimate \citep[e.g.,][]{1979ApJ...233L..57T, 1989A&A...210..114C, 2002MNRAS.329...29C, 2003ApJ...585..443S}; (iii) the absolute luminosity of Type I X‑ray bursts, assuming that the peak luminosity is equal to the Eddington luminosity \citep[e.g.,][]{1993MNRAS.262..511E, 2017ApJ...836..111K, 2020ApJS..249...32G}, but the result is highly dependent on assumptions about the elemental composition and the mass of NS, making this method less reliable. Distances to UCXBs are primarily derived from (i) and (iii), since their companions are low‑mass WDs or helium stars with extremely faint optical emission, making it difficult to determine distances through (ii) \citep[e.g.,][]{2003MNRAS.344..201J, 2009A&A...508..297D, 2010ApJ...714..894H}. Distances to other LMXBs shown in Figure \ref{fig:3} are derived by (ii) \citep{2008NewAR..51..733N}, which also have good estimates.

It should be noted that the $ \dot{M} $ calculated in Equation (\ref{eq:2}) represents the long‑term average values. Owing to disk instability \citep{2008A&A...486..523L}, the actual accretion rates may exhibit substantial fluctuations, such as the outburst seen in M15 X‑2 \citep{2025A&A...695A..44D}. However, assuming that GR consistently dominates the angular momentum loss, we argue that the long‑term average accretion rate is close to the theoretical prediction. There are also some uncertainties in $ \overline{L}_{\rm X} $. For example, 4U 1850-087 has only a few observations, and M15 X‑2 has an outburst luminosity about an order of magnitude higher than its normal luminosity \citep{2025A&A...695A..44D}. These factors can lead to uncertainties of up to an order of magnitude in the average luminosity. However, according to Figure \ref{fig:2}, even with such large uncertainties, the radiative efficiency of NS systems is still significantly higher than that of BH systems.

Deriving an unabsorbed X-ray flux from the observed data involves several steps, which can be summarized as the following six steps. 

(1) Select suitable data based on the characteristics of the sources. The five UCXBs considered in this work are all located in GCs, where the source density can be high, so imaging X-ray instruments are particularly important for resolving individual sources, especially for M15 X-2 \citep{2005MNRAS.357..325H} and 47 Tuc X-9 \citep{2017MNRAS.467.2199B}. M15 X-2 and 47 Tuc X-9 are faint and lie in crowded regions containing sources of comparable brightness. Therefore, high-spatial-resolution \textit{Chandra} observations are essential for these two sources because \textit{Chandra} provides sub-arcsecond on-axis imaging. For 4U 1820-303, which is significantly brighter than nearby sources, observations with non-imaging instruments such as \textit{NICER} can also be used \citep{2023MNRAS.525.2366M}. For 4U 0513-40, \textit{Chandra} observations reveal a pulsar within $ \sim 10 $ arcsec, but it is too faint to significantly affect the source analysis \citep{2011MNRAS.414L..41F}. For 4U 1850-087, although no high-resolution \textit{Chandra} imaging analysis is available, its host cluster NGC 6712 is relatively diffuse, with a core radius of $ \sim56 $ arcsec \citep{1996MNRAS.282L..37H}. Therefore, for 4U 0513-40 and 4U 1850-087, we also consider X-ray data from other observatories.

(2) Extract source and background events from the calibrated event data for imaging instruments, which mainly includes \textit{Chandra} and \textit{Swift}/XRT. The source region is generally selected based on the point-spread function (PSF), the source brightness, the source crowding, the off-axis angle, and possible pile-up. For \textit{Chandra}/ACIS, the on-axis 90\% encircled-energy radius is approximately 2.0 arcsec at 1.49 keV and 2.5 arcsec at 6.4 keV, see \url{https://cxc.harvard.edu/proposer/POG/html/chap6.html}. For example, \cite{2017MNRAS.467.2199B} used a 1.8-arcsec source-extraction radius and a background annulus extending from 2.7 to 7.4 arcsec for 47 Tuc X-9, illustrating that the extraction region can be smaller than the radius enclosing 90\% of the PSF in order to reduce contamination from nearby sources. For \textit{Swift}/XRT LSXPS catalogue, the source detection procedure uses PSF fitting rather than a single fixed aperture. The catalogue provides the parameter \texttt{PSF\_Fit\_Radius}, defined as the radius of the circular region over which the PSF fit is performed, in XRT image pixels ($1$ pixel $=2.357$ arcsec). The radius depends on the source signal-to-noise ratio and the 1SXPS/2SXPS catalogues use radii of approximately 12-30 pixels \citep{2020ApJS..247...54E}, while the upper limit of the background extraction radius is 12 pixels (\url{https://www.swift.ac.uk/2SXPS/uldocs.php}).

(3) For non-imaging instruments, such as \textit{NICER}, a conventional source extraction radius is not defined. Instead, source and background spectra are obtained using event selection and a background-modeling procedure, such as the \texttt{nibackgen3C50} tool \citep{2023MNRAS.525.2366M}. Further details are provided in the \textit{NICER} Mission Guide: \url{https://heasarc.gsfc.nasa.gov/docs/nicer/mission_guide/}.

(4) Generating the instrument response files, including the redistribution matrix file (RMF) and ancillary response file (ARF). RMF describes the probability that a photon of a given energy is detected in a particular detector channel, thereby accounting for the detector's finite energy resolution and energy redistribution. ARF describes the energy-dependent effective collecting area and relevant efficiency corrections, including effects such as detector efficiency, mirror effective area, vignetting, and PSF or aperture corrections. Together, the RMF and ARF fold a physical spectral model into detector space, allowing the predicted count spectrum to be compared with the observed spectrum. In general, the appropriate response is observation and extraction dependent, particularly for the ARF; response files should therefore be generated or selected using the calibration database (CALDB) appropriate to the observation.

(5) Fitting an appropriate spectral model. Commonly used continuum models for X-ray binaries include power law (\texttt{powerlaw}) and standard thin multi-colour disc blackbody (\texttt{diskbb}), sometimes with additional components such as \texttt{bbodyrad} and \texttt{thcomp} when required by the data \citep[e.g.][]{2001ApJ...561L.101W, 2005ApJ...627..926J, 2017MNRAS.467.2199B, 2023MNRAS.525.2366M}. The derived unabsorbed flux depends on both the adopted spectral model and the neutral hydrogen column density, $N_{\rm H}$, because the absorption is significant at low X-ray energies \citep{2000ApJ...542..914W}. For the sources considered here, $ N_{\rm H} $ is generally below $ 10^{22} \ \mathrm{cm}^{-2} $, and they are not classified as supersoft sources. Therefore, for the relatively hard or moderately soft spectra relevant to most of these UCXBs, the uncertainty due to the choice of fitting model is typically less than 50\% \citep{2017MNRAS.466.1019S}.

(6) For \textit{Chandra}/HRC, spectral fitting cannot be usefully carried out. Although HRC provides excellent spatial resolution ($ ~0.9 $ arcsec), its intrinsic non-dispersive energy resolution is too poor for useful spectral fitting, see \url{https://cxc.harvard.edu/proposer/POG/html/chap7.html}. Thus, for an HRC observation without a grating, one generally cannot derive an unabsorbed flux by fitting an X-ray spectrum in the same way as for ACIS. Instead, the measured HRC count rate is converted to an energy flux by assuming a spectral model and an absorbing column density, and then folding this assumed model through the HRC instrumental response.

\begin{figure*}[!t]
    \gridline{\fig{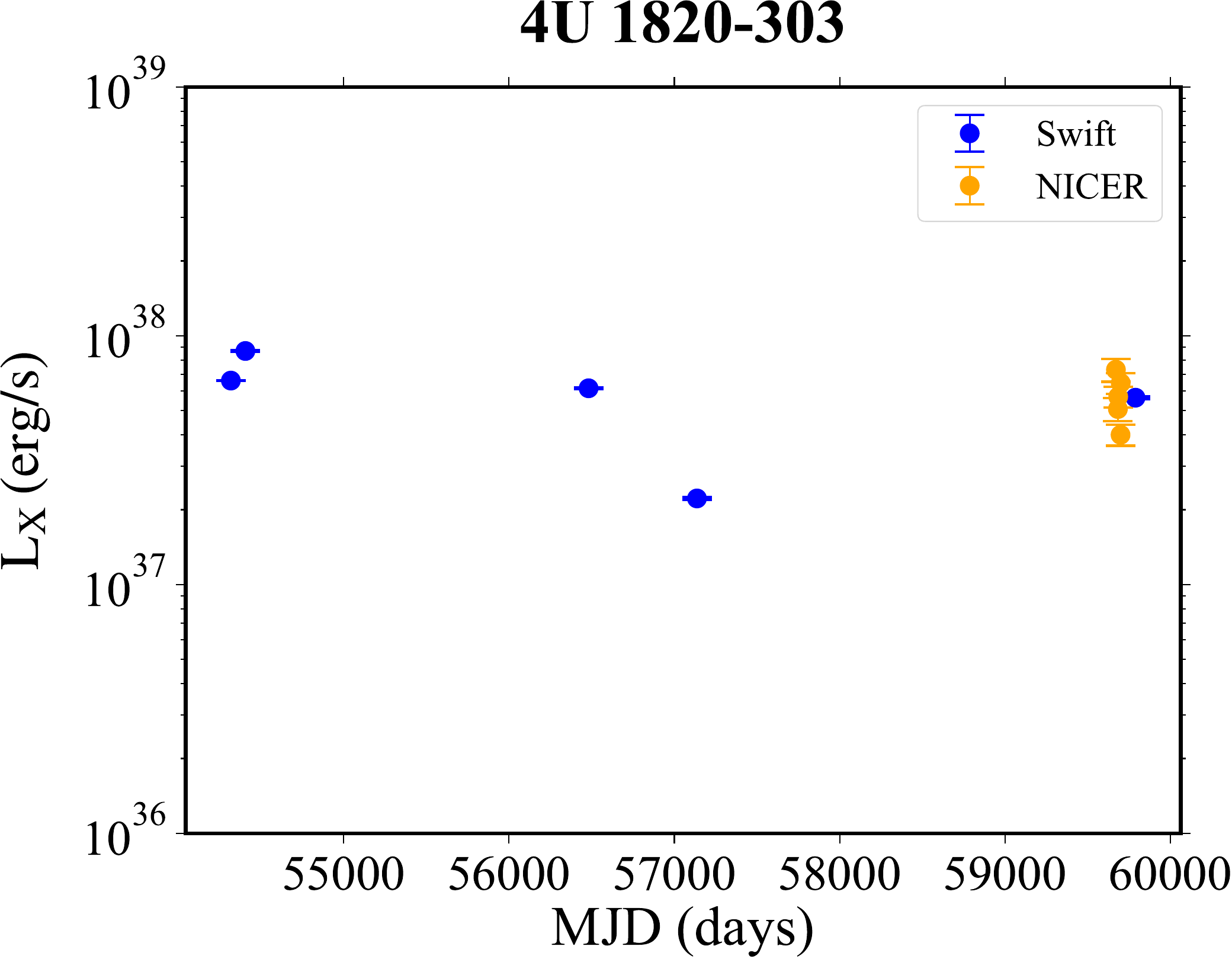}{0.40\textwidth}{(a)}
              \fig{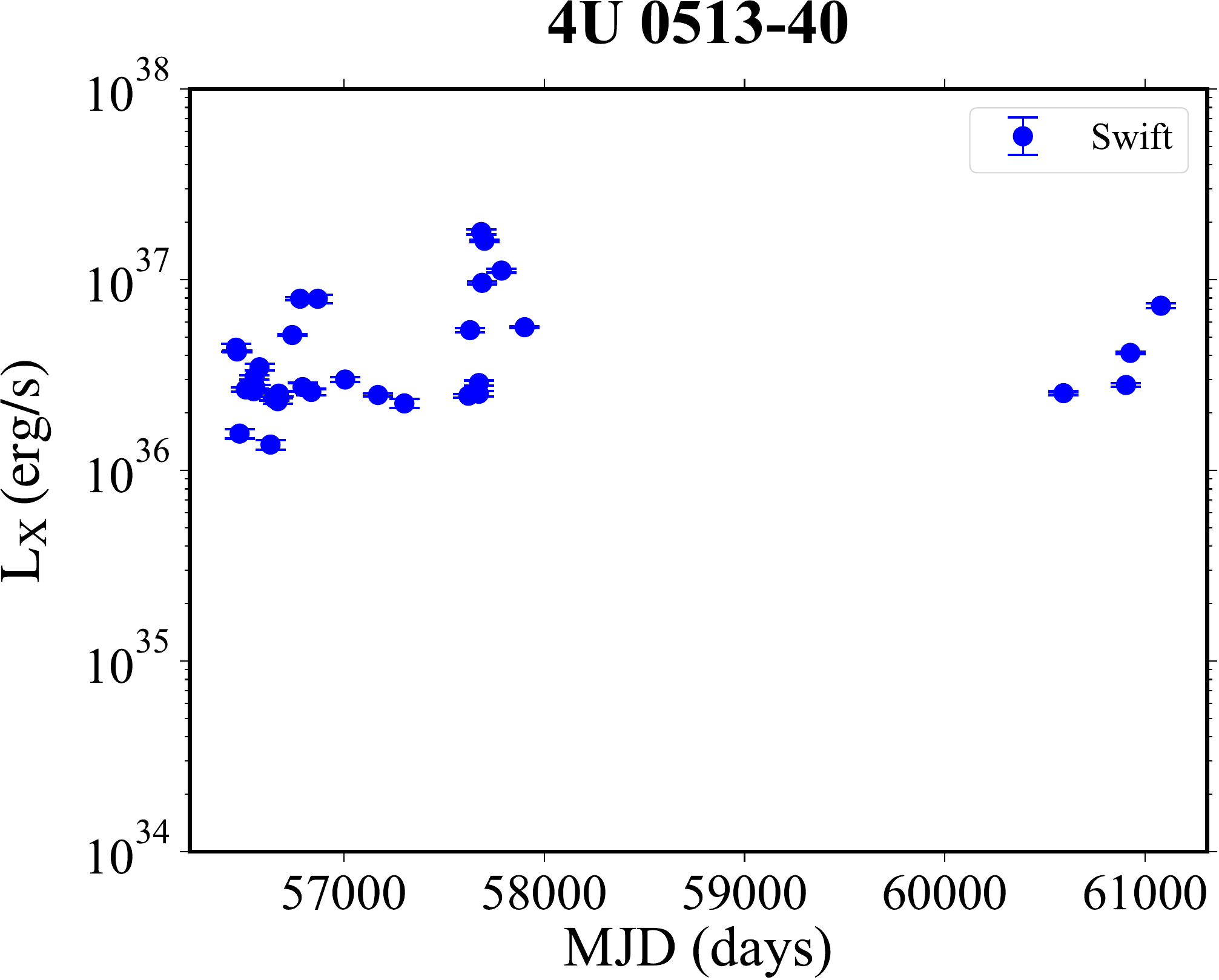}{0.40\textwidth}{(b)}}
    \gridline{\fig{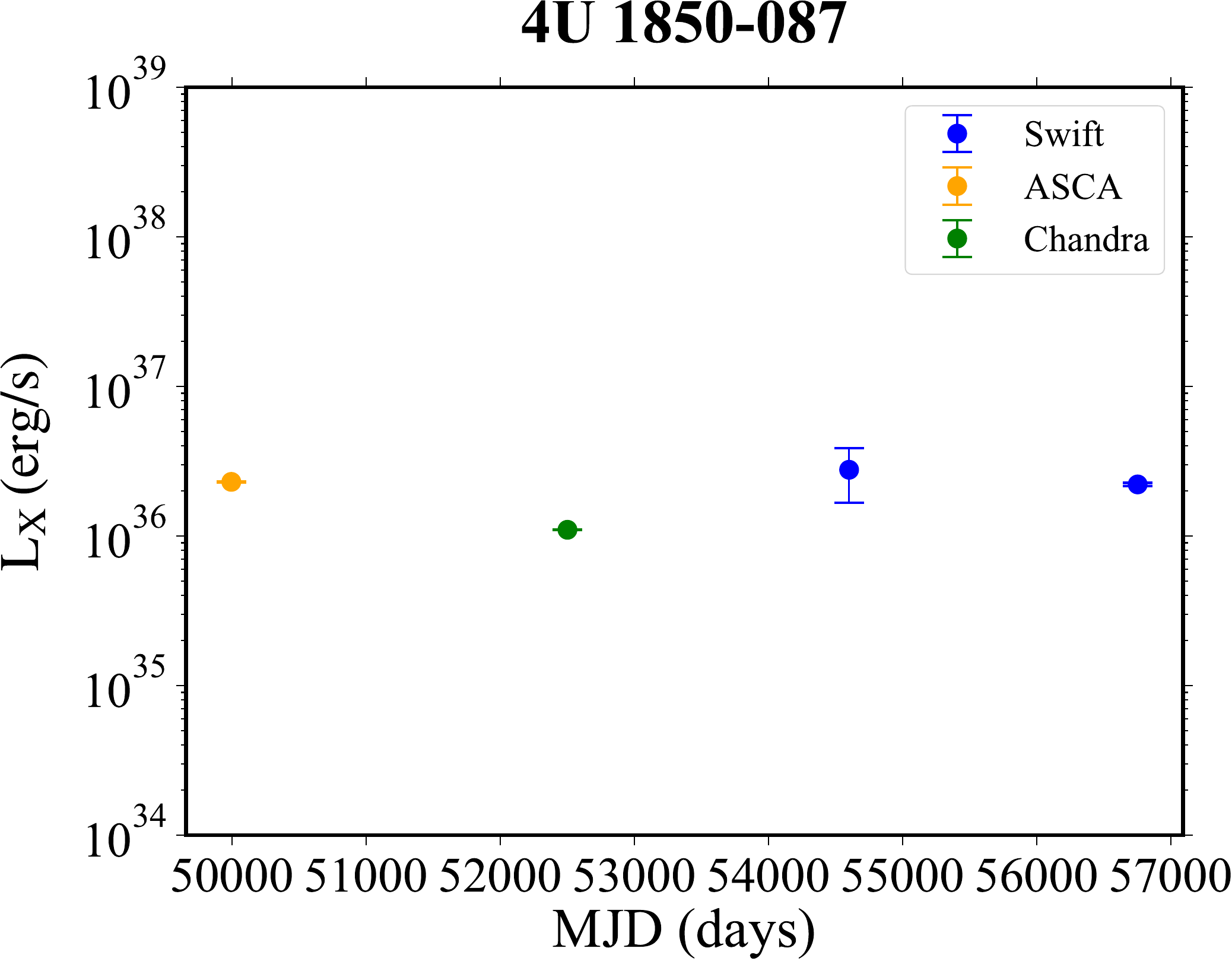}{0.40\textwidth}{(c)}
              \fig{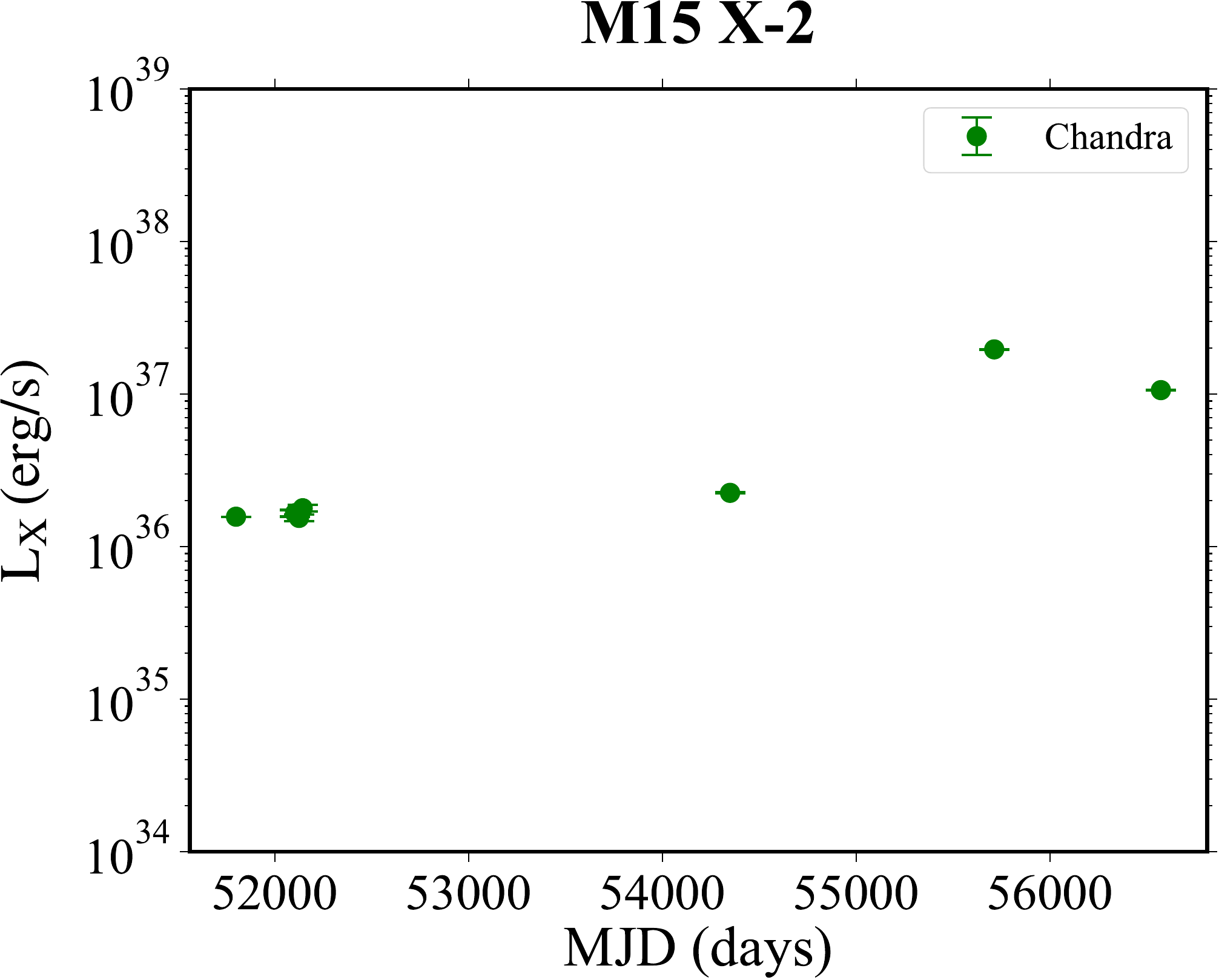}{0.40\textwidth}{(d)}}
    \gridline{\fig{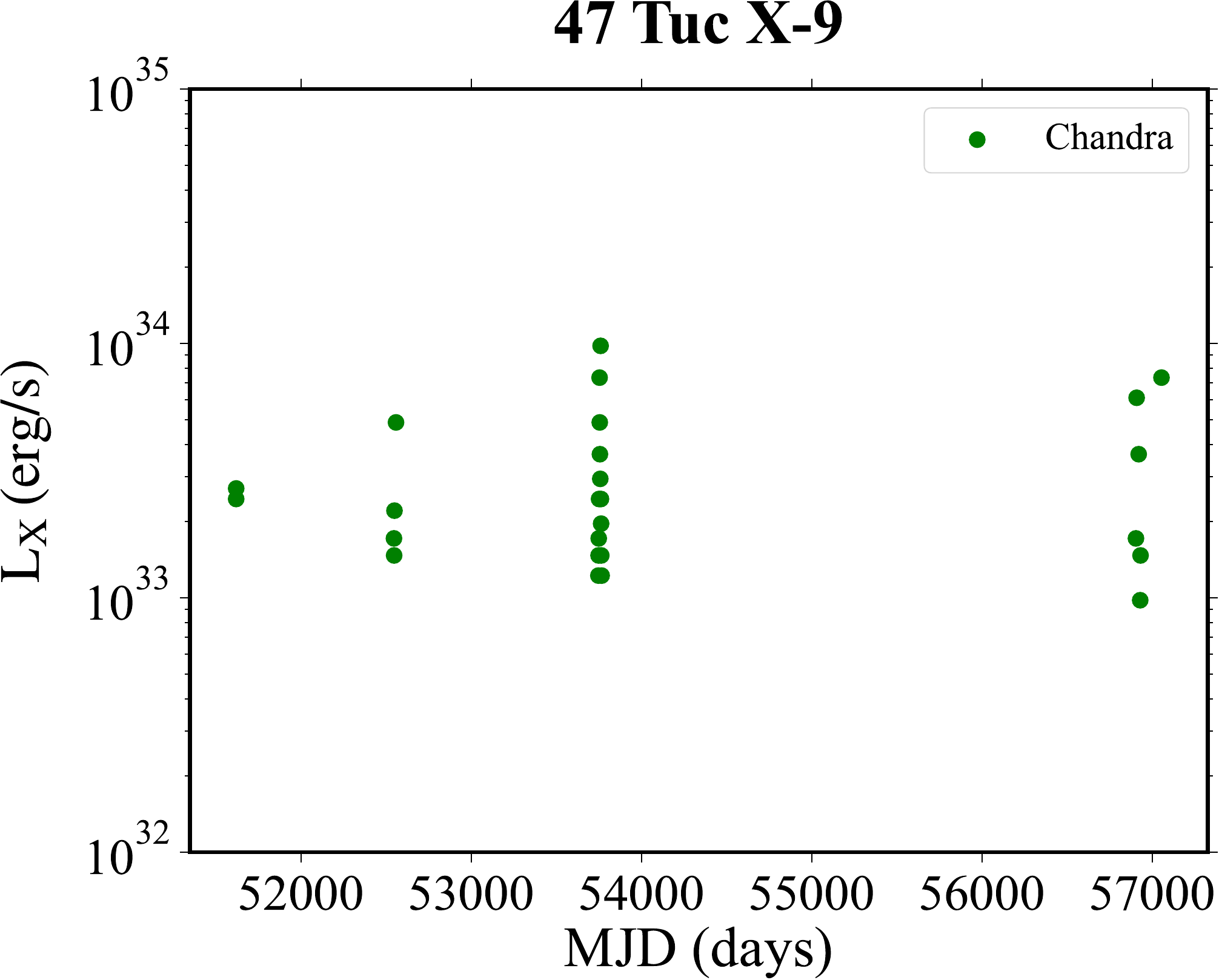}{0.40\textwidth}{(e)}}
    
    \caption{Light curves of the selected UCXBs. The X-ray energy band is 0.5-10 keV, and some error bars are not visible because they are smaller than the symbols. (a) 4U 1820-303: \textit{Swift} and \textit{NICER} \citep{2023MNRAS.525.2366M} data are used. \textit{Swift} data are derived from the LSXPS catalog. (b) 4U 0513-40: \textit{Swift} data are also derived from the LSXPS catalog. (c) 4U 1850-087: \textit{Swift} (from the LSXPS catalog and \cite{2008ATel.1551....1P}), \textit{ASCA} \citep{2001ApJ...560L..59J}, and \textit{Chandra} data \citep{2005ApJ...627..926J} are used. (d) M15 X-2: A nearby source with comparable brightness is present; therefore, only \textit{Chandra} data \citep{2001ApJ...561L.101W, 2005MNRAS.357..325H, 2011ATel.3393....1S, 2013ATel.5490....1P} are used to avoid source confusion. (e) 47 Tuc X-9: Nearby sources are also present; therefore, only \textit{Chandra} data \citep{2017MNRAS.467.2199B} are used.} 
    \label{fig:4}
\end{figure*}

Finally, RMF and ARF calibration files are continuously updated as the instrument calibration improves. Consequently, response files generated with different software or CALDB versions may not be identical, but their effect on our results is negligible. Therefore, we directly adopt the flux results from historical literature and catalogues. The light curves of the five sources are shown in Figure \ref{fig:4}. The detailed observational information is presented in Tables~\ref{tab:2}, \ref{tab:3}, \ref{tab:4}, \ref{tab:5}, and \ref{tab:6}, which include the observatory/instrument (Obs./Inst.), MJD, exposure, X-ray luminosity ($ L_{\rm X} $), X-ray luminosity uncertainty ($ \delta L_{\rm X} $), and the references. The \textit{Swift} LSXPS Catalogue cited in the references is available online at \url{https://www.swift.ac.uk/LSXPS/}, and $ \delta L_{\rm X} $ for the \textit{Swift} is from the $ 1\sigma $ statistical uncertainty of the count rate. Errors for the other data are discussed in the references shown in Tables~\ref{tab:2} to \ref{tab:6}. Except for \textit{Swift}/XRT observations (0.3–10 keV), all X‑ray luminosities presented in the following tables are in the 0.5–10 keV band. In this work, we assume that $ L_{\rm 0.3-10 \: keV} \approx L_{\rm 0.5-10 \: keV} $.

\clearpage
\clearpage

\begin{deluxetable*}{llllll}[h]
\tablenum{2}
\tablewidth{0pt}
\tablecaption{Observational data for 4U 1820-303\label{tab:2}}
\tablehead{
\colhead{Obs./Inst.} & \colhead{MJD} & \colhead{Exposure (s)} & \colhead{$ L_{\rm X} \: (\rm erg\,s^{-1}) $} & \colhead{$ \delta L_{\rm X} \: (\rm erg\,s^{-1}) $} & \colhead{References} \\
}
\startdata
\textit{Swift}/XRT & 54318 & 9472 & $ 6.6 \times 10^{37} $ & $ 2.6 \times 10^{35} $ & \textit{Swift} LSXPS Catalogue \\
 & 54406 & 14784 & $ 8.7 \times 10^{37} $ & $ 4.7 \times 10^{35} $ & \\
 & 56482 & 66499 & $ 6.2 \times 10^{37} $ & $ 3.7 \times 10^{35} $ & \\
 & 57137 & 3035 & $ 2.2 \times 10^{37} $ & $ 2.5 \times 10^{35} $ & \\
 & 59790 & 374 & $ 5.6 \times 10^{37} $ & $ 6.6 \times 10^{35} $ & \\
\hline
\textit{NICER} & 59671 & 8100 & $ 7.3 \times 10^{37} $ & $ 7.7 \times 10^{36} $ & \cite{2023MNRAS.525.2366M} \\
 & 59683 & 4200 & $ 5.1 \times 10^{37} $ & $ 5.4 \times 10^{36} $ & \\
 & 59685 & 1100 & $ 5.7 \times 10^{37} $ & $ 5.4 \times 10^{36} $ & \\
 & 59699 & 2600 & $ 4.0 \times 10^{37} $ & $ 3.8 \times 10^{36} $ & \\
 & 59700 & 6500 & $ 6.5 \times 10^{37} $ & $ 6.2 \times 10^{36} $ & \\
\enddata
\end{deluxetable*}

\startlongtable
\begin{deluxetable*}{llllll}
\tablenum{3}
\tablewidth{0pt}
\tablecaption{Observational data for 4U 0513-40\label{tab:3}}
\tablehead{
\colhead{Obs./Inst.} & \colhead{MJD} & \colhead{Exposure (s)} & \colhead{$ L_{\rm X} \: (\rm erg\,s^{-1}) $} & \colhead{$ \delta L_{\rm X} \: (\rm erg\,s^{-1}) $} & \colhead{References} \\
}
\startdata
\textit{Swift}/XRT & 56463 & 92    & $4.4 \times 10^{36}$  & $ 2.2 \times 10^{35}$ & \textit{Swift} LSXPS Catalogue \\
 & 56468 & 49225 & $4.2 \times 10^{36}$  & $ 4.8 \times 10^{34}$ & \\
 & 56480 & 103   & $1.6 \times 10^{36}$  & $ 9.0 \times 10^{34}$ & \\
 & 56511 & 415   & $2.7 \times 10^{36}$  & $ 7.3 \times 10^{34}$ & \\
 & 56551 & 392   & $2.6 \times 10^{36}$  & $ 7.3 \times 10^{34}$ & \\
 & 56555 & 401   & $3.1 \times 10^{36}$  & $ 8.4 \times 10^{34}$ & \\
 & 56559 & 43414 & $2.7 \times 10^{36}$  & $ 6.9 \times 10^{34}$ & \\
 & 56580 & 5901  & $3.5 \times 10^{36}$  & $ 1.4 \times 10^{35}$ & \\
 & 56635 & 118   & $1.4 \times 10^{36}$  & $ 8.0 \times 10^{34}$ & \\
 & 56654 & 738   & $2.4 \times 10^{36}$  & $ 4.8 \times 10^{34}$ & \\
 & 56671 & 321   & $2.3 \times 10^{36}$  & $ 7.5 \times 10^{34}$ & \\
 & 56676 & 412   & $2.5 \times 10^{36}$  & $ 7.8 \times 10^{34}$ & \\
 & 56743 & 14647 & $5.1 \times 10^{36}$  & $ 7.1 \times 10^{34}$ & \\
 & 56782 & 22577 & $7.9 \times 10^{36}$  & $ 1.4 \times 10^{35}$ & \\
 & 56795 & 113   & $2.7 \times 10^{36}$  & $ 1.4 \times 10^{35}$ & \\
 & 56838 & 224   & $2.6 \times 10^{36}$  & $ 9.7 \times 10^{34}$ & \\
 & 56871 & 58    & $7.9 \times 10^{36}$  & $ 4.0 \times 10^{35}$ & \\
 & 57007 & 527   & $3.0 \times 10^{36}$  & $ 7.9 \times 10^{34}$ & \\
 & 57171 & 3131  & $2.5 \times 10^{36}$  & $ 4.7 \times 10^{34}$ & \\
 & 57303 & 105   & $2.2 \times 10^{36}$  & $ 1.2 \times 10^{35}$ & \\
 & 57622 & 26268 & $2.5 \times 10^{36}$  & $ 5.1 \times 10^{34}$ & \\
 & 57631 & 554   & $5.4 \times 10^{36}$  & $ 1.3 \times 10^{35}$ & \\
 & 57675 & 8828  & $2.5 \times 10^{36}$  & $ 8.1 \times 10^{34}$ & \\
 & 57675 & 5876  & $2.9 \times 10^{36}$  & $ 9.2 \times 10^{34}$ & \\
 & 57687 & 242   & $1.8 \times 10^{37}$  & $ 5.4 \times 10^{35}$ & \\
 & 57691 & 72007 & $9.6 \times 10^{36}$  & $ 1.7 \times 10^{35}$ & \\
 & 57703 & 729   & $1.6 \times 10^{37}$  & $ 2.4 \times 10^{35}$ & \\
 & 57788 & 376   & $1.1 \times 10^{37}$  & $ 2.8 \times 10^{35}$ & \\
 & 57903 & 34636 & $5.6 \times 10^{36}$  & $ 7.1 \times 10^{34}$ & \\
 & 60592 & 690   & $2.5 \times 10^{36}$  & $ 5.5 \times 10^{34}$ & \\
 & 60904 & 2659  & $2.8 \times 10^{36}$  & $ 6.5 \times 10^{34}$ & \\
 & 60925 & 3148  & $4.1 \times 10^{36}$  & $ 6.4 \times 10^{34}$ & \\
 & 61078 & 238   & $7.3 \times 10^{36}$  & $ 2.2 \times 10^{35}$ & \\
\enddata
\end{deluxetable*}

\begin{deluxetable*}{llllll}[h]
\tablenum{4}
\tablewidth{0pt}
\tablecaption{Observational data for 4U 1850-087\label{tab:4}}
\tablehead{
\colhead{Obs./Inst.} & \colhead{MJD} & \colhead{Exposure (s)} & \colhead{$ L_{\rm X} \: (\rm erg\,s^{-1}) $} & \colhead{$ \delta L_{\rm X} \: (\rm erg\,s^{-1}) $} & \colhead{References} \\
}
\startdata
\textit{Swift}/XRT & 54598-54602 & 8200 & $ 2.8 \times 10^{36} $ & $ 1.1 \times 10^{36} $ & \textit{Swift} LSXPS Catalogue \\
 & 56754 & 2080 & $ 2.2 \times 10^{36} $ & $ 5.4 \times 10^{34} $ &  \\
\hline
\textit{Chandra}/ACIS & 52502 & 50000 & $ 1.1 \times 10^{36} $ & - & \cite{2005ApJ...627..926J} \\
\hline
\textit{ASCA} & 49996 & 20000 & $ 2.3 \times 10^{36} $ & $ 2.3 \times 10^{34} $ & \cite{2001ApJ...560L..59J} \\
\enddata
\end{deluxetable*}

\begin{deluxetable*}{llllll}[h]
\tablenum{5}
\tablewidth{0pt}
\tablecaption{Observational data for M15 X-2\label{tab:5}}
\tablehead{
\colhead{Obs./Inst.} & \colhead{MJD} & \colhead{Exposure (s)} & \colhead{$ L_{\rm X} \: (\rm erg\,s^{-1}) $} & \colhead{$ \delta L_{\rm X} \: (\rm erg\,s^{-1}) $} & \colhead{References} \\
}
\startdata
\textit{Chandra}/ACIS & 51800 & 21000 & $ 1.6 \times 10^{36} $ & - & \cite{2001ApJ...561L.101W} \\
\hline
\textit{Chandra}/HRC & 52104 & 10619 & $ 1.7 \times 10^{36} $ & $ 8.3 \times 10^{34} $ & \cite{2005MNRAS.357..325H} \\
 & 52125 & 10753 & $ 1.5 \times 10^{36} $ & $ 7.7 \times 10^{34} $ & \\
 & 52144 & 12372 & $ 1.8 \times 10^{36} $ & $ 8.9 \times 10^{34} $ & \\
\hline
\textit{Chandra}/HRC & 54349 & 2140 & $ 2.3 \times 10^{36} $ & $ 2.3 \times 10^{34} $ & \cite{2011ATel.3393....1S} \\
 & 55712 & 1500 & $ 2.0 \times 10^{37} $ & $ 5.9 \times 10^{34} $ & \\
\hline
\textit{Chandra}/HRC & 56572 & 15100 & $ 1.1 \times 10^{37} $ & - & \cite{2013ATel.5490....1P} \\
\enddata
\end{deluxetable*}

\startlongtable
\begin{deluxetable*}{llllll}
\tablenum{6}
\tablewidth{0pt}
\tablecaption{Observational data for 47 Tuc X-9\label{tab:6}}
\tablehead{
\colhead{Obs./Inst.} & \colhead{MJD} & \colhead{Exposure (s)} & \colhead{$ L_{\rm X} \: (\rm erg\,s^{-1}) $} & \colhead{$ \delta L_{\rm X} \: (\rm erg\,s^{-1}) $} & \colhead{References} \\
}
\startdata
\textit{Chandra}/ACIS & 51619 & 31700 & $2.7 \times 10^{33}$ & - & \cite{2017MNRAS.467.2199B} \\
 & 51619 & 31700 & $2.4 \times 10^{33}$ & - & \\
 & 52546 & 65240 & $1.7 \times 10^{33}$ & - & \\
 & 52547 & 65340 & $1.5 \times 10^{33}$ & - & \\
 & 52549 & 65400 & $2.2 \times 10^{33}$ & - & \\
 & 52558 & 68770 & $4.9 \times 10^{33}$ & - & \\
 & 56905 & 40880 & $1.7 \times 10^{33}$ & - & \\
 & 56909 & 50040 & $6.1 \times 10^{33}$ & - & \\
 & 56921 & 24700 & $3.7 \times 10^{33}$ & - & \\
 & 56930 & 9130  & $9.8 \times 10^{32}$ & - & \\
 & 56932 & 16240 & $1.5 \times 10^{33}$ & - & \\
 & 57055 & 40280 & $7.3 \times 10^{33}$ & - & \\
\hline
\textit{Chandra}/HRC & 53747 & 50000 & $1.2 \times 10^{33}$ & - & \cite{2017MNRAS.467.2199B} \\
 & 53748 & 51000 & $1.5 \times 10^{33}$ & - & \\
 & 53749 & 50000 & $1.7 \times 10^{33}$ & - & \\
 & 53751 & 52000 & $2.4 \times 10^{33}$ & - & \\
 & 53754 & 48000 & $7.3 \times 10^{33}$ & - & \\
 & 53755 & 45000 & $4.9 \times 10^{33}$ & - & \\
 & 53756 & 47000 & $3.7 \times 10^{33}$ & - & \\
 & 53758 & 44000 & $2.9 \times 10^{33}$ & - & \\
 & 53760 & 97000 & $9.8 \times 10^{33}$ & - & \\
 & 53762 & 50000 & $2.4 \times 10^{33}$ & - & \\
 & 53763 & 52000 & $2.0 \times 10^{33}$ & - & \\
 & 53764 & 50000 & $1.5 \times 10^{33}$ & - & \\
 & 53766 & 48000 & $1.2 \times 10^{33}$ & - & \\
\enddata
\end{deluxetable*}

\end{document}